\documentclass{article}
\usepackage[left=3cm,right=3cm,top=3cm,bottom=3cm]{geometry} 
\usepackage{amsmath} 
\usepackage{authblk}

\usepackage{epstopdf}
\usepackage{subfigure}
\usepackage{graphicx,hyperref}
\usepackage{amssymb}
\usepackage{amsmath}
\usepackage{multirow}
\usepackage{float}
\usepackage{mathtools}
\mathtoolsset{showonlyrefs=true}
\usepackage{subfigure}
\usepackage{xcolor}

\begin{document}


\title{\textit{Hidden Markov model for discrete circular-linear wind data time series } }


\author[1]{Gianluca Mastrantonio}
\author[2]{Gianfranco Calise}

\affil[1]{ Roma Tre University, Via Silvio D'Amico 77,  Rome, 00145, Italy}
\affil[2]{Department of Earth Science, University of Rome``Sapienza''}

\date{}

\maketitle

\begin{abstract}
In this work, we deal with a bivariate time series of wind speed and direction. Our observed data have  peculiar features, such as informative missing values, non-reliable  measures under a specific condition and interval-censored data, that we take into account in the model specification.

We analyze the time series with a non-parametric Bayesian hidden Markov model, introducing  a new emission distribution based on the invariant wrapped Poisson, the Poisson and the hurdle  density, suitable to model our data.    The model is estimated on simulated datasets and on the real data example that motivated this work.

\end{abstract}

\begin{keywords}
Invariant wrapped Poisson; hurdle model; discrete circular variable; non-parametric Bayesian; Dirichlet process.
\end{keywords}

%

\section{Introduction}

The analysis of time series of  meteorological data  has an increasing interest in many fields. Their analysis is interesting  in order to have a better understanding of the atmospheric phenomena, to determine the climate of a geographical data or to predict the occurrence of extreme events.
In this work, motivated by our real data example, we are interested in  the modelling of a bivariate time series of wind speed and direction.

The analysis of wind time series  have been carried out  by means of different approaches, for example   mixture type models (\cite{Ailliot2012}), harmonic analyses  (\cite{belu2013}),  ARMA-GARCH models (\cite{Lojowska2010}) or stochastic differential equations (\cite{Zarate2013}).  Among the others, the hidden Markov model (HMM), a  class of mixture model for  time series,   proved to  be well suited to model wind data,  see for example
\cite{Martin1999}, \cite{Holzmann2006},  \cite{Ailliot2012}, \cite{Bulla2012}, \cite{lagona2015} or \cite{mastrantonio2015}.
In the HMM,   the data-generative distribution  is  expressed 
as a mixture, with components that belong, generally, to an ``easy tractable'' distribution (called \emph{emission distribution} or \emph{regime-specific density}). The belonging to a component of the mixture, called also \emph{regime} or \emph{state}, depends on an underlying and unobservable discrete valued Markov process.

The wind direction is a circular variable, i.e. a variable that represents an angle or a   point over the unit circle, while the speed is a linear (or inline) one. Hence, if a time series of  wind direction and speed is modelled through an HMM, the emission distribution must be defined over a mixed  circular-linear domain. Generally conditional independence between the circular and linear variables is  assumed, but exceptions exist (see for example \cite{Bulla2012,mastrantonio2015}).

Although HMMs for wind data have been proposed in the literature, here, due to particular features of our data, a direct application of the models previously proposed is not possible. The measured linear  and circular variables are interval-censored (\cite{Lindsey1998}) and they  are recorded as discrete variables; more precisely the wind speed is recorded as integer knots  and the  direction is measured on a discrete scale with  36 equally spaced  values over the unit circle. Furthermore, the direction is strictly related to the wind speed as when the latter is too small, the direction is not recorded and, due to the low instrument sensitivity,  a recorded value of 0 or 1   wind speed  is not reliable.

To take into account all these  features,  we introduce a new bivariate distribution to be used as the regime-specific density  in the HMM.   The model is estimated under a non-parametric Bayesian framework, using the hierarchical Dirichlet  process-HMM (HDP-HMM) of \cite{Teh2006} and the modification introduced by \cite{fox2011}, namely the sticky HDP-HMM (sHPD-HMM). The sHPD-HMM  allows us to estimate, as a model parameter, the number of regimes occupied by the time series.  The model is applied to simulated and real data examples.

The paper is organized as follows. In Section  \ref{sec:motivo} we describe   the motivating example while in Section \ref{sec:dist} we introduce the  circular-linear distribution by first formalizing the marginal distribution of the linear component, Section \ref{sec:lin}, and then the one of the circular, conditioned to the linear one, Section \ref{sec:circ}. Section \ref{sec:model} is devoted to the model specification while Section \ref{sec:prior} gives implementation details. Section \ref{sec:ex} contains the simulated examples (Section \ref{sec:sim})  and the real data one (Section \ref{sec:real}).  The paper ends with a discussion (Section  \ref{sec:disc}).

\section{Motivating example}\label{sec:motivo}

In last decades the Italian cost suffers an intensification of erosion, in particular in South Italy, along the coastline on the Tyrrhenian side, with manmade coastal structures. This is obviously a result of both natural and anthropogenic causes. The study of wind time series, recorded along the cost, is of a great importance since the wind  generates  waves, a significant factor of coastal geomorphology.

The data  are  recorded with an anemometer, located on the rocky cape of Capo Palinuro, in the town of Centola,  province of Salerno, South Italy.  The meteorological station of Capo Palinuro  is one of the coastal stations managed by the Meteorological Service of the Military Italian Air Force.  The instrument is placed away from obstacles, 10 meters above  ground. 
The data are
provided by the National Center of Aeronautical Meteorology and Climatology (C.N.M.C.A.), special office of the Meteorological Service of the Italian Air Force.

In this work we focus on the year 2010, that  is a particular one. While the station, for decades, has recorded  winds blowing between North and North-West or  South and South-East octants, in 2010 the prevailing winds (the directions with the higher frequency) are the ones blowing from North-East and South-West, which is almost the opposite of the general trend. Then, it is important to understand and characterize in a greater details, i.e. through an HMM, the wind  distribution of this year. 

%

%
%
\subsection{Data description}

The  linear variable  is interval-censored, i.e. it is recorded  with  an error of  1 knot,  and  let $Y^*$ be the recorded wind speed, if $Y^*=c$, with $c \in \mathbb{Z}^+$,  the real (continuous) variable has value $\geq \min\{c-0.5,0 \}$ but $<c+0.5$.   Further  uncertainty is added to the  values below 2 knots as the instrument sensitivity does not allow reliable recordings. 

Also the wind direction is interval-censored and it is measured in degrees 
on a discrete scale with 36 distinct values.  If the recorded circular variable, $X$, assumes value $c$, with $c \in \{ \frac{2 \pi}{36}j\}_{j=0}^{35}$\footnote{The direction recorded by the instrument assumes values in $(0,10,20,\dots, 350)$ but we rescale it to   $[0,2 \pi)$. }, the real direction has value $\geq c-\frac{1}{2}\frac{2 \pi}{36}$ but $<c+\frac{1}{2}\frac{2 \pi}{36}$.  
If the (continuous) wind speed is too low, i.e. close to 0, the instrument could not be able to record  $X$ and we have a missing  recording.
This missing gives  information about the non-reliable linear recording, i.e.  they are \emph{informative}  and  \emph{non-ignorable}, (\cite{RUBIN1976}) and then
the  process that generates the missing observation  must be taken into account in the model specification.

\begin{figure}[t!]
	\centering
	{\subfigure[Wind speed]{\includegraphics[scale=0.4]{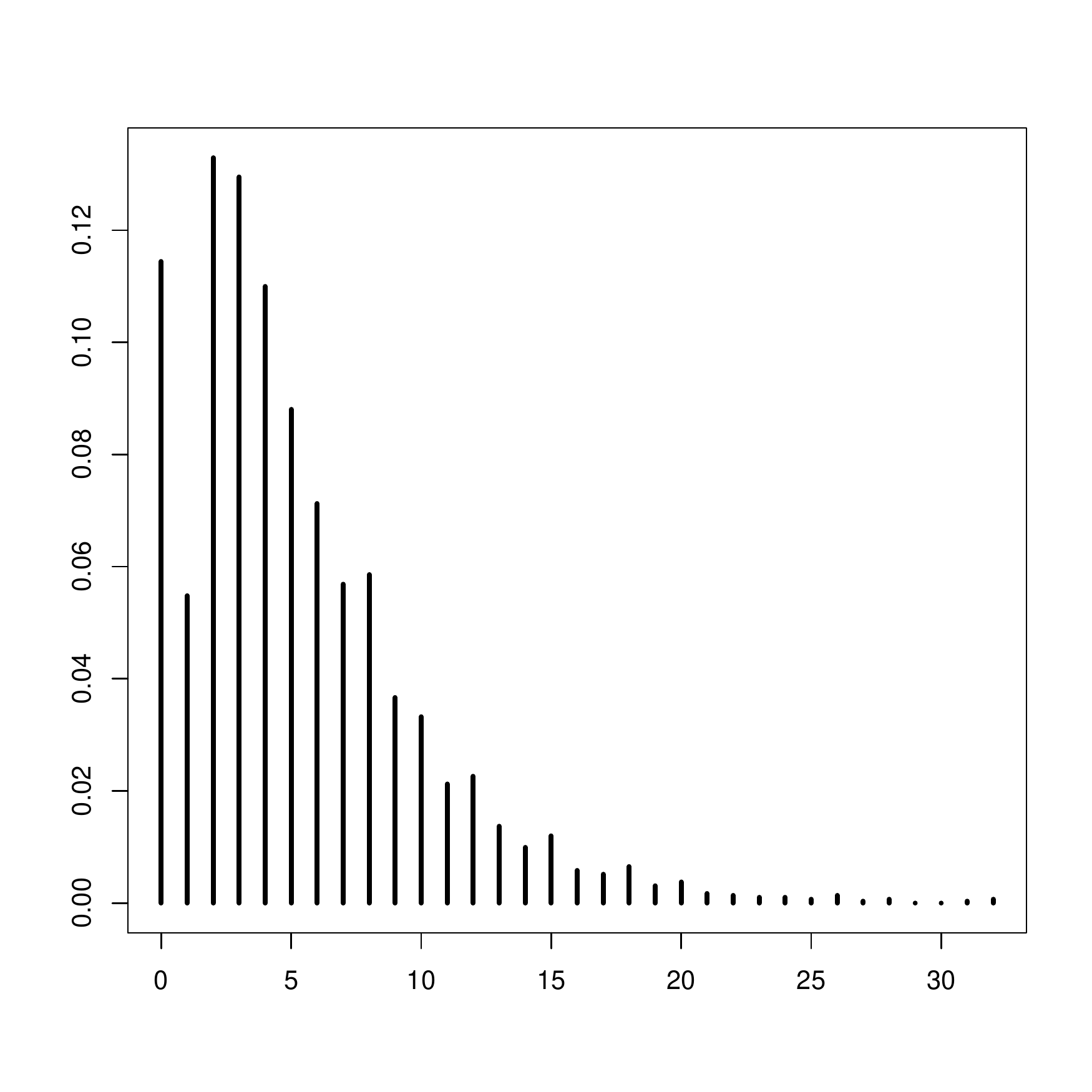}}}
	{\subfigure[Wind direction]{\includegraphics[scale=0.4]{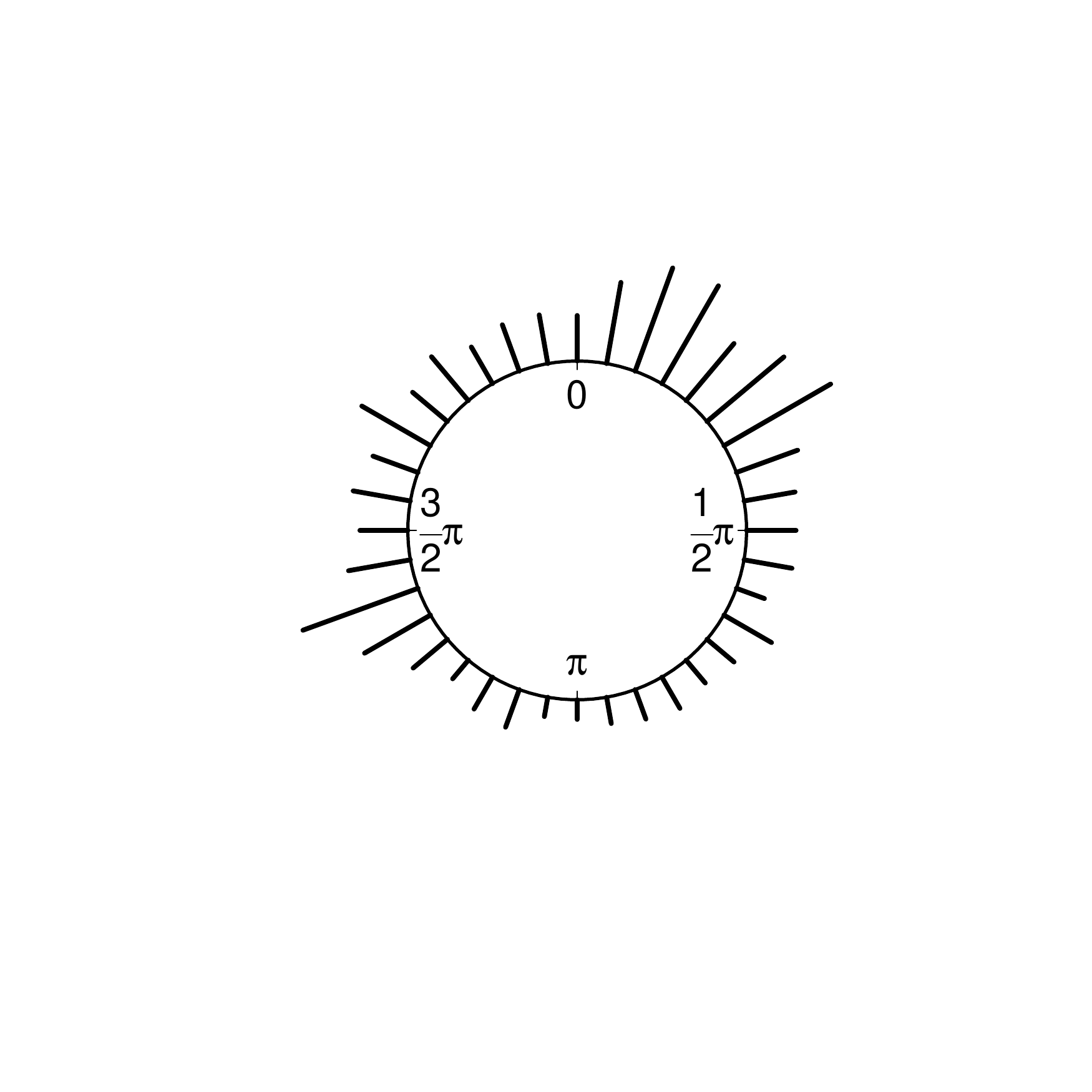}}}
	\caption{Linear (a) and circular (b) barplot of the variable used in the  real data example.} \label{fig:obs}
\end{figure}

Due to the instrument malfunction,  we have  2920 observations with, respectively, 3  and 2 non-informative missing  in the circular and linear variables. $y_t^*$ assumes value 0  494 times while
 $x_t $ is equal to $ \{\emptyset \}$   328 times.   Figure \ref{fig:obs} shows
barplots of the circular and linear data.


\section{The model } \label{sec:dist}

Before the formalization  of the HMM (Section \ref{sec:model}), we introduce the circular-linear distribution we  use in the model specification.

\subsection{The linear distribution} \label{sec:lin}

We code 
the information carried by the recorded linear variable ($Y^*$) into two new variables. One representing the ``true'' discrete wind speed  with support $\mathbb{Z}^+$, $Y$, and a   Bernoulli variable $W$,  that assumes value  0 when $Y^*<2$ and  1 otherwise. The marginal distribution of $Y$ is setted to be Poisson with parameter $\lambda_y$: $Y\sim P(\lambda)$.  Between $Y$ and $Y^*$  the following relation exists:
\begin{equation} \label{eqy11}
\Big\{
\begin{array}{lccc}
Y = Y^* &  \mbox{ } & \mbox{if }Y^*\geq 2, \\
Y \in \{0,1\}  & \mbox{ } & \mbox{if }Y^*<2. 
\end{array}
\end{equation}
\\
Let  $W$ be a binary random variable such that
\begin{align} 	
W=1 & \mbox{ if } Y^*\ge 2,\label{eq:w1}\\ 
W=0 & \mbox{ if } Y^* <2 ,\label{eq:w2}
\end{align}
with  
\begin{equation}\label{eq:ber}
W \sim Bern(1-e^{-\lambda_{y}}(1+\lambda_{y})),
\end{equation}
 where  $Bern(\cdot)$  indicates the Bernoulli distribution.  We define the distribution of $Y|W, \lambda_y$ as follows:
\begin{equation}  \label{eq:y1}
P(y|w=1, \lambda_y)= 
\frac{\lambda_y^{y}e^{-\lambda_y}}{y!}   \frac{1}{ 1-e^{-\lambda_{y}}(1+\lambda_{y}) }  I(y \in {\mathbb{Z}^+\backslash \{0,1\}}) 
\end{equation}
and 
\begin{equation} \label{eq:y2}
P(y|w=0, \lambda_y)= 
\left(\frac{\lambda_{y}e^{-\lambda_y}}{e^{-\lambda_{y}}(1+\lambda_{y})}\right)^y \left(1-\frac{\lambda_{y}e^{-\lambda_y}}{e^{-\lambda_{y}}(1+\lambda_{y})}\right)^{1-y}  I(y \in {\{0,1\}})
\end{equation}
where $I(y \in {A})$ is the indicator function.

\subsection{The circular distribution} \label{sec:circ}

To model properly the non-ignorable missing value of $X$, we specify the domain of the circular variable  as a mixed one,  composed of a discrete circle $\mathbb{D} = \{\frac{2 \pi}{36}j\}_{j=0}^{l-1}$, with $36$ equally spaced values, and  the empty set, $\{ \emptyset\}$. The direction assumes value $\{ \emptyset\}$ if the measure is a missing non-ignorable  and 
then the distribution of the circular variable is an hurdle one. The hurdle distribution, first introduced by \cite{Mullahy1986}, assumes that the  observation, in this case the wind direction, comes from two data-generative  processes.  With probability $\nu^*$ the observation belongs  to the hurdle, i.e.   the empty set,  otherwise  it belongs to the   other portion of the mixed domain, i.e. $\mathbb{D}$, and a distribution over $\mathbb{D}$ must be chosen.
%
%
%
%
%
%

Since the circular variable can assume value $\{ \emptyset\}$ only if the discrete wind speed is 0, the  probability $\nu^*$ must depend on $Y$.  Indeed  $(X=\{\emptyset\}) \Rightarrow (Y=0)$   but it is not true that $(Y=0)\Rightarrow (X=\{\emptyset\})$, as $X$ may be measured when a light wind  is blowing (below 1 knot). We specify the hurdle probability as  $\nu^*= \nu I(y=0)$ and, as required,  $\nu^*\neq 0$  only if $Y=0$.\\
In the literature several distributions  for discrete circular variables have been proposed, see for example the wrapped Binomial  (\cite{Girija2014}) or the wrapped Weibull  (\cite{Sarma2011}).    \cite{mastrantonio2015f} proved that for most of them, 
the inference they allow  strongly depends on the  choice of the reference system origin  and  orientation, while a  ``proper'' distribution must be independent on these choices, for details see \cite{mastrantonio2015f}.   As distribution over $\mathbb{D}$,   we choose the invariant wrapped Poisson (IWP), proposed by \cite{mastrantonio2015f}, since it is  a proper and flexible distribution for discrete circular data that is  easy to implement in a Bayesian framework.  

%

The density of $X$ is
\begin{equation} \label{eq:hiwp1}
P(x|y, \lambda_{x},\eta,\xi) = \left( \nu^*   \right)^{I(x=\emptyset)} \left(   (1-\nu^* )   P(x|x \in \mathbb{D},\lambda_{x},\eta,\xi)   \right)^{I(x\in\mathbb{D})}.
\end{equation}
where $P(x|x \in \mathbb{D},\lambda_{x},\eta,\xi)$ is the pmf of the IWP:
\begin{equation} \label{eq:unw3}
P(x|x \in \mathbb{D}, \lambda_{x},\eta, \xi)= \sum_{k=0  }^{\infty} \frac{\lambda^{(\eta \theta- \xi   ) \text{ mod } (2 \pi) 36/(2 \pi)+ k36 }e^{-\lambda}}{((\eta \theta- \xi   )\text{ mod } (2 \pi) 36/(2 \pi)+ k36)!}, \, \xi  \in \mathbb{D},\, \eta \in \{-1,1\}.  
\end{equation}
 \eqref{eq:unw3} is obtained by wrapping the density of a linear transformation of a   Poisson distributed variable. More in details, we first  obtain the density of $Q^* = \eta\left(  Q \frac{2 \pi}{36}  +\xi \right)$, where  $Q \sim P(\lambda_x)$, that is 
\begin{equation} \label{eq:yty}
 \frac{\lambda^{(\eta q^*- \xi   )36/(2 \pi) }e^{-\lambda}}{((\eta q^*- \xi   )36/(2 \pi))!}, 
\end{equation}
then we find the distribution of the associated wrapped variable $X = Q^* \text{ mod } 2 \pi$,  that is \eqref{eq:unw3}. Note that the transformation $X = Q^* \text{ mod } 2 \pi$  wraps the linear variable $Q^*$, and its density,  around the discrete circle $\mathbb{D}$. $k$, i.e.  \emph{winding number}, identify in \eqref{eq:unw3} the $k^{th}$ portion of the domain of $Q^*$ that is wrapped around the  circle.


We can compute in closed form the directional mean and  concentration of the IWP (\cite{mastrantonio2015f}), i.e. the circular counterparts of linear  mean and concentration:
\begin{align}
\mu =& \eta\xi+\lambda \sin\left(\eta \frac{2 \pi}{36}\right), \label{eq:cricmean}\\
c =& e^{-\lambda \left(1- \cos \left(\frac{2 \pi}{36}\right)\right)}.\label{eq:cricconc}
\end{align}
Equations \eqref{eq:cricmean} and \eqref{eq:cricconc}, as the mean and concentration for the inline variables, are useful statistics  to describe the circular variable behaviour.

It is not easy to work directly with equation \eqref{eq:unw3},   since it involves an  infinite sum.   When a wrapped  distribution  is used (\cite{coles98,Jona2013,mastrantonio2015c}),  a standard approach is to introduce the latent random variable  $K$,  and to  work with the joint density of $(X,K)$ that is the summand in \eqref{eq:unw3} and does not require the infinite sum evaluation:
\begin{equation}  \label{eq:aa}
P(x,k|x \in \mathbb{D}, \lambda_{x},\eta, \xi)=  \frac{\lambda^{(\eta \theta- \xi   ) \text{ mod } (2 \pi) 36/(2 \pi)+ kl }e^{-\lambda}}{((\eta \theta- \xi   )\text{ mod } (2 \pi) 36/(2 \pi)+ kl)!}.
\end{equation}
The winding number is needed also to define an efficient sampling scheme for the IWP parameters, see \cite{mastrantonio2015f} and Section \ref{sec:prior}. \\
The augmented hurdle type density is
\begin{equation} \label{eq:hiwp2}
P(x,k|y, \lambda_{x},\eta,\xi) = \left( \nu^*   \right)^{I(x=\emptyset)} \left(   (1-\nu^* )   P(x,k|x \in \mathbb{D},\lambda_{x},\eta,\xi)   \right)^{I(x\in\mathbb{D})}.
\end{equation}

We write $X,K,Y,W|\lambda_{x},\eta, \xi, \nu,\lambda_{y} \sim HiwpP_{\mathbb{D}}(\lambda_x, \eta, \xi,\nu,\lambda_y)$ if  $W|\lambda_y $ is distributed as  \eqref{eq:ber},   $Y|W, \lambda_y$ as in \eqref{eq:y1} and \eqref{eq:y2} and $X,K|Y, \lambda_{x},\eta,\xi$
as in  \eqref{eq:hiwp2}.

\subsection{The HMM specification} \label{sec:model}

%

Following the formalization introduced in Sections \ref{sec:lin} and \ref{sec:circ}, we  proceed by modeling the four-variate time series $\{ \mathbf{x}, \mathbf{k}, \mathbf{y},\mathbf{w} \}$, where $\mathbf{x}= \{ x_t\}_{t=1}^T$, $\mathbf{k}=\{k_t\}_{t=1}^T$, $\mathbf{y}= \{y_t\}_{t=1}^T$ and  $\mathbf{w}= \{w_t\}_{t=1}^T$.
We time-cluster the data with a   non-parametric Bayesian HMM, namely the sHDP-HMM, that allows us to group the time series in homogeneous regimes, as the standard HMM, but it does not need to assume known the number of regimes $R$ that can be  estimated along with the other model parameters. \\
The belonging to a regime  is coded via a discrete random variable $z_t$; if $z_t=r$, at time $t$ the system is in regime $r$. In the sHDP-HMM is assumed that $z_t \in \mathbb{Z}^+ \backslash \{0 \} $, i.e. the number of regimes is potentially infinite. Indeed since $T$ is finite, the elements of the time series $\mathbf{z}=\{z_t \in \mathbb{Z}^+ \backslash  \{0 \} \}_{t=1}^T$ will assume only a finite number of states ($R$),  i.e. the number of non-empty regimes.

The sHDP-HMM is a hierarchical model where,  let $\boldsymbol{\psi}_r=\{\lambda_{x,r}, \eta_r, \xi_r, \nu_r,\lambda_{y,r}\}$ be the vector of parameters,  the first stage is
\begin{align}
P( \mathbf{x}, \mathbf{k}, \mathbf{y}, \mathbf{w}|\mathbf{z},  \{\boldsymbol{\psi}_{r}\}_{r=1}^{\infty})  =&   \prod_{t =1}^T \prod_{r=1}^{\infty} \left[  P(x_t,k_t,y_t,w_t|\boldsymbol{\psi}_{r})  \right]^{I({r=z_t})}, \label{eq:hmm1}\\
X_t,K_t,Y_t,W_t|\boldsymbol{\psi}_r  \sim & HiwpP_{\mathbb{D}}(\lambda_{x,r}, \eta_k, \xi_k,\nu_k,\lambda_{y,r}), \label{eq:hmm12}
\end{align}
At the second level of the hierarchy, $\mathbf{z}$ is assumed to follow a first order Markov process, i.e. $P(z_t|z_{t-1},z_{t-2},\dots , z_{1})= P(z_t|z_{t-1})$, with 
\begin{equation} \label{eq:hmm2}
P(z_t | z_{t-1} , \boldsymbol{\pi}_{z_{t-1}}) = \pi_{z_{t-1}z_t}.
\end{equation}
and $\boldsymbol{\pi}_j= \{ \pi_{ji} \}_{i=1}^{\infty}$ is   the $j^{th}$ row of the transition matrix, i.e. the matrix that rules the probabilities to move from one state to another.  The initial state $z_0$, and the associated  vector of probabilities, cannot be estimated consistently since we have no observations  at time 0 (for details see \cite{cappe2005}) and, without loss of generality, we set $z_0=1$.  The equations \eqref{eq:hmm1}, \eqref{eq:hmm12} and \eqref{eq:hmm2} define an HMM with an infinite number of states, i.e. $z_t \in \mathbb{Z}^+\backslash \{0\}$, infinite number of parameters, $\{\boldsymbol{\psi}_{r}\}_{r=1}^{\infty}$, and vectors of probabilities of infinite length, $\boldsymbol{\pi}_j$s.

The sHDP-HMM specification is concluded assuming the following: 
\begin{align}
\boldsymbol{\pi}_{{r}}  | \rho,\gamma,\{\beta_j \}_{j=1}^{\infty},\boldsymbol{\psi}_{r}& \sim DP\left(\gamma,  \sum_{j=1}^{\infty}  \left(  (1-\rho) \beta_j+\rho I(r=j)    \right)  \right), \,  \rho\in [0,1], \gamma\in \mathbb{R}^+, \label{eq:distpi}\\
\beta_r &=  \beta_r^* \prod_{j=1}^{r-1}(1-\beta_j^*)  \\
\beta_r^* |\tau&\sim  B(1,\tau),  \, \tau \in \mathbb{R}^+\\
\boldsymbol{\psi}_{r}| H &\sim H ,  \label{eq:H}
\end{align}
where $\beta_r>0, \,r=1,2,\dots, \infty $  and  $\sum_{r=1}^{\infty} \beta_{r}=1$, i.e. $\{\beta_r \}_{r=1}^{\infty}$ is a vector of probabilities,  $B(\cdot,\cdot)$ is the Beta distribution,  $DP(\cdot)$   indicates   the Dirichlet process  while  $H$ is a distribution over  $\boldsymbol{\psi}_r$ that acts as a prior for the model parameters. 
Without loss of generality, let suppose that the first $R$ states are the non-empty ones, and let $\boldsymbol{\pi}_r^*=\left( \pi_{r1},\dots , \pi_{rR}, \sum_{j=R+1}^{\infty}\pi_{rj} \right)$. The definition of the Dirichlet process (see for example \cite{fox2011})  implies  that  \eqref{eq:distpi} can be written as 
\begin{align}
&\boldsymbol{\pi}_r^* |\rho,\gamma,\{\beta_j \}_{j=1}^{\infty},\boldsymbol{\psi}_{r}\sim \\ 
&Dir\left(\gamma((1-\rho)\beta_{1}+\rho I(r=1))      ,\dots,\gamma((1-\rho)\beta_{R}+\rho I(r=R) )    , \gamma(1-\rho)\sum_{j=R+1}^{\infty}\beta_j  \right) \label{eq:sw},
\end{align}
where $Dir(\cdot)$ is the Dirichlet distribution.

From  \eqref{eq:sw} it is clear that, for $r,j=1,\dots,R$, $E(\pi_{rj}) = (1-\rho)\beta_r+\rho I(r=j)$,  $Var(\pi_{rj}) = \frac{ ((1-\rho)\beta_{j}+\rho I(r=j)) (1- (1-\rho)\beta_{j}-\rho I(r=j))  }{\gamma+1}$. The vector $\{\beta_r \}_{r=1}^{\infty}$ and $\rho$ rules the mean value of the $\boldsymbol{\pi}_r^*$ (and then also the one of $\boldsymbol{\pi}_r$) and  $\rho $ is an additional  weight added to  the auto-transition probability (or self-transition probability) $\pi_{rr}$. $\rho$  is needed otherwise the HMM tends to create redundant states  (\cite{teh2010}).
$\gamma$ is directly proportional to  the precision (the inverse of the variance) of  $\pi_{rj}$. The parameters $\rho$, $\gamma$ and $\tau$  rule the number of non empty regimes, $R$, and as they increase, $R$ decreases (see \cite{fox2011}).

For  a  more detailed explanation of the properties and interpretation of the sHDP-HMM we refer  the reader to \cite{fox2011}.

\subsection{Prior distributions and implementation details} \label{sec:prior}

To specify the prior distribution over $\boldsymbol{\psi}_r$, $H$, we follow two standard advices. The first is to  have prior distributions that make the model parameters easy to update in the Markov chain Monte Carlo (MCMC) algorithm, i.e. within a Gibbs step. The second advice is to use prior distributions that allow an easy learning from the data and a robust estimation of the posterior distribution and it is usually achieved using weakly informative priors.

In the simulated and real data examples (Section \ref{sec:ex}), we will use the same set of priors. Specifically  $\lambda_{y,r}\sim G(a_y,b_y)I(0,c_y)$, where $G(\cdot,\cdot)I(0,\cdot)$ indicates the truncated gamma distribution expressed in terms of shape and rate that, under specific set of parameters, can resemble the uniform distribution over $[0,c_y]$. Due to the standard conjugacy between the Poisson and the gamma, the prior on $\lambda_{y,r}$  leads to a truncated gamma full conditional. We use a uniform distribution for $\nu_r$ that leads to a beta  full conditional.
For the IWP parameters we follows the work of \cite{mastrantonio2015f}. They show that  an efficient MCMC algorithm can be defined if we set an upper bound on the range of  $\lambda_{x,r}$, i.e. $\lambda_{x,r}\in [0,\lambda_{x,max}]$, we use a truncated gamma distribution for $\lambda_{x,r}$  and we fix the maximum value that $k_t$s can assume. The value $\lambda_{x,max}$ is chosen  such that  the IWP evaluated with $\lambda_{x,r}=\lambda_{x,max}$ is  indistinguishable from a circular discrete uniform.  \cite{mastrantonio2015f} propose to find the value $\lambda_{x,max}$ by using the wrapped normal approximation of the IWP and the truncation strategy of  \cite{Jona2013}.  Once we have the maximum value of $\lambda$, we  find the maximum value of the $k_t$s that is needed to obtain a reasonable approximation of the IWP, i.e. $k_{max}= \lceil \frac{3 \sqrt{\lambda_{max}}}{l}+\frac{\lambda_{max}}{l}-\frac{1}{2} \rceil$, where $\lceil \cdot \rceil$ is the ceiling operator.  Then following \cite{mastrantonio2015f}, regardless  on the prior distributions for $\eta_{r}$ and $\xi_r$, we can define and efficient Gibbs sampler. For  $\eta_r$ and $\xi_r$ we use uniform distributions  in their respective domains. \\
$R$ depends on $\rho$, $\gamma$ and $\tau$  and then we choose to threat them as random quantities, with non informative prior for each of them, so  to be non informative on the distribution of $R$.  Following \cite{fox2011}, the following priors,  $\rho \in U(0,1)$, $\gamma \sim G(a_{\rho},b_{\rho})$  and $\tau \sim G(a_{\tau},b_{\tau})$, lead to full conditionals easy to simulate. The MCMC update of the regime indicator variables is done with the beam sampler  (\cite{VanGael2008}).\\
With regard to the time series $\mathbf{y}$,  we can have two types of missing observations; when  $y_t^*=0,1$, i.e. an informative missing  since  we know that the value of $y_t$ is either 0 or 1, and when  $y_t^*$ is missing due to the malfunction of the anemometer, i.e. a non informative missing. In both cases  the missing observations are estimated during the model fitting. If $y_t$ is missing and $y_t^*=0$ then  $w_t=0$ and the full conditional of $y_t$ is      $Bern\left( \frac{\lambda_{y,r}e^{-\lambda_{y,r}}}{e^{-\lambda_{{y,r}}}(1+\lambda_{y,r})} \right)$  if $x_t \in \mathbb{D}$, see equation \eqref{eq:yty}, while $y_t=0$ if $x = \{\emptyset\}$. If  both $y_t$ and $y_t^*$ are missing, then   $w_t$ is missing as well and, if $x = \{ \emptyset \}$ we have that $y_t=0$ and $w_t=0$, otherwise if $x \in \mathbb{D}$, we can simulate from the joint full conditional of $(y_t,w_t)$ by first simulate $y_t$ from a Poisson with parameter $\lambda_{y,r}$, and then we  set $w_t=1$ if $y_t \geq 2$ and 0 otherwise. 
The non informative missing observation in the time series $\mathbf{x}$, due to a malfunction of the station, can be easily simulated, i.e. if $y_t \neq 0$ we simulate a value  from the IWP, while if $y_t =0$  we simulate $w_t$ from  a $Bern(\nu_r)$ and, if  $w_t=1$ then $x_t = \{ \emptyset \}$ while if $w_t=0$ we simulate a value from the IWP.

\begin{figure}[t!]
	\centering
	{\subfigure[Linear densities]{\includegraphics[scale=0.4]{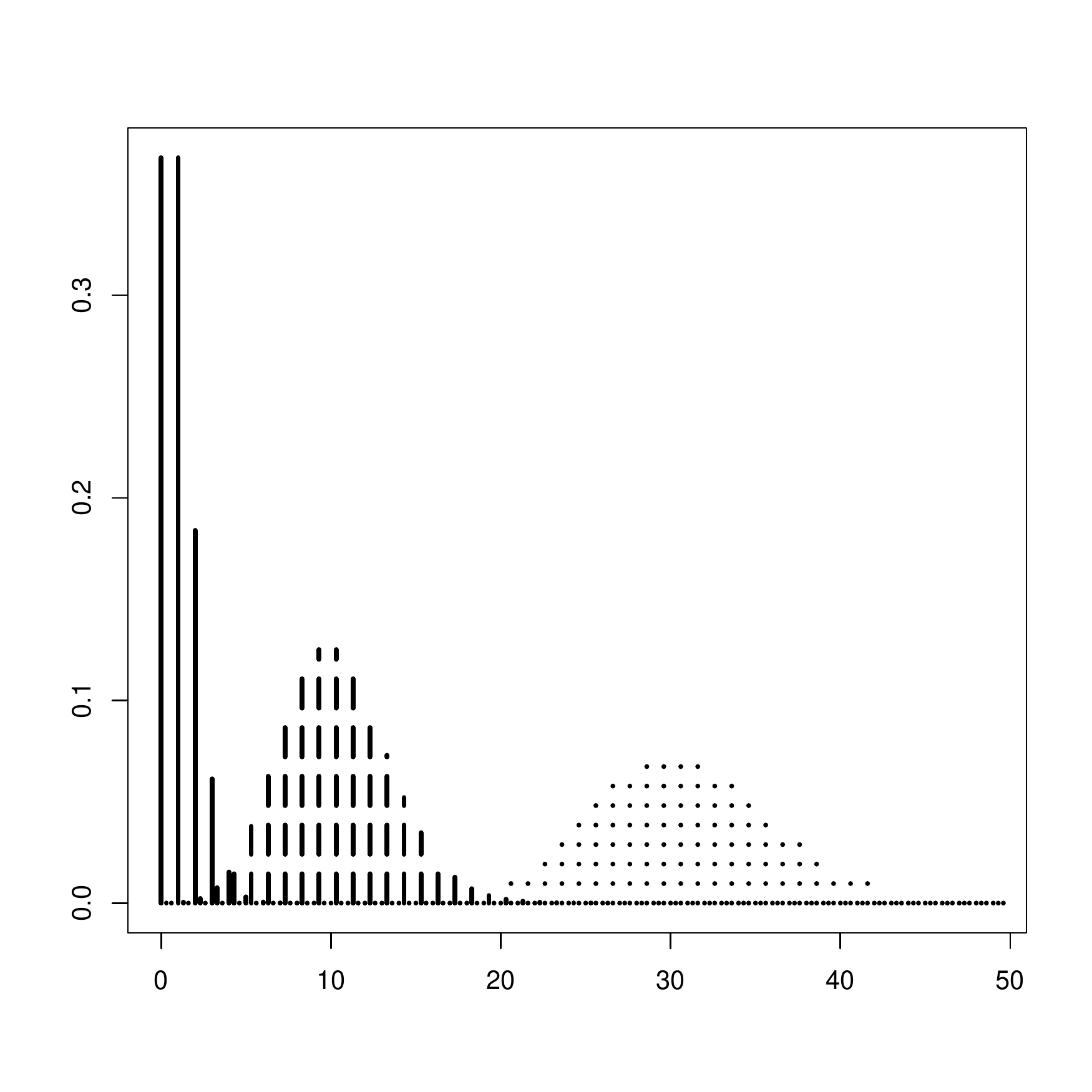}}}
	{\subfigure[Circular densities]{\includegraphics[scale=0.4]{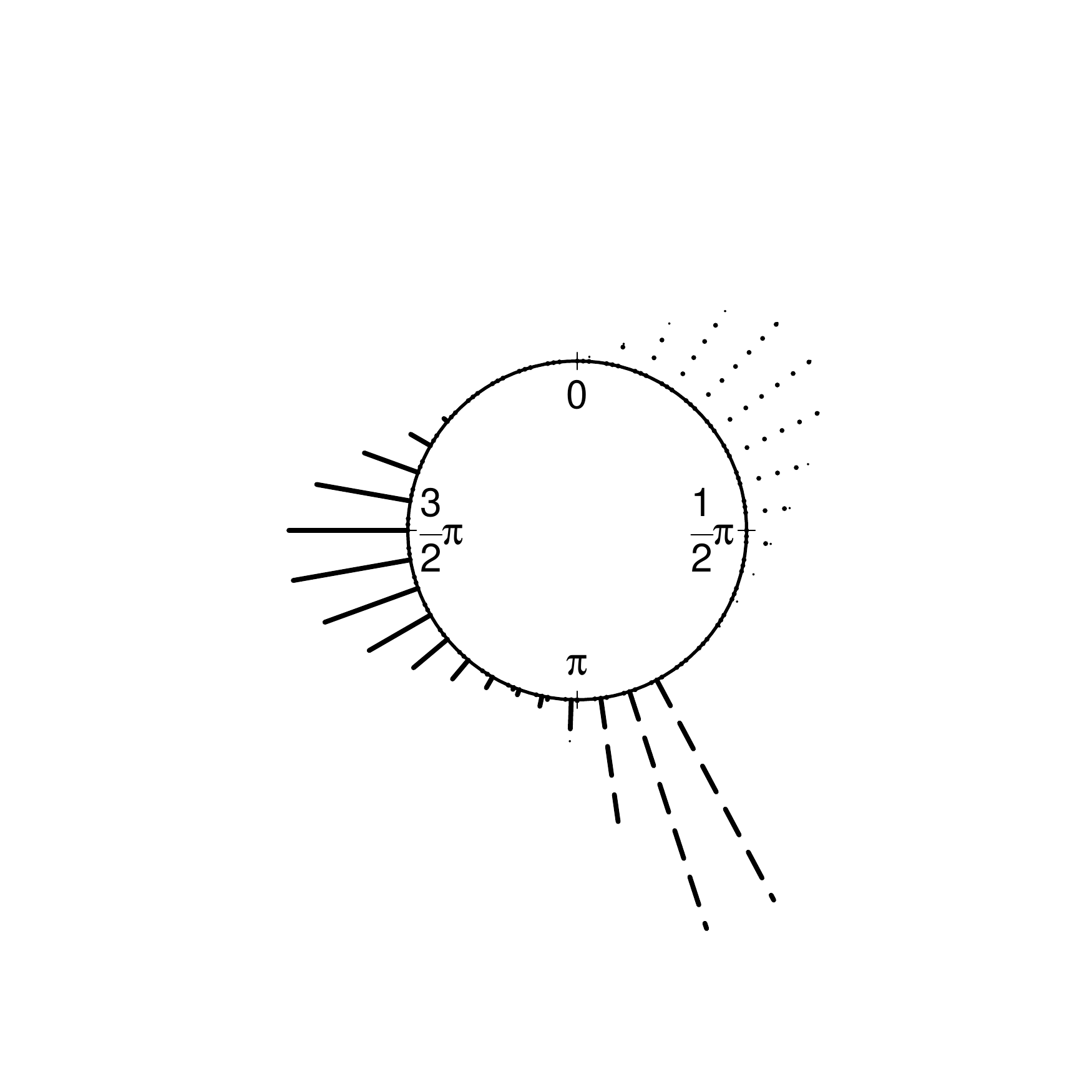}}}
	\caption{Simulate example: densities used to generate the example 1. The solid line is the first regime, the dashed the second and the dotted the third } \label{fig:Ex1}
\end{figure}
\begin{figure}[t!]
	\centering
	{\subfigure[Linear densities]{\includegraphics[scale=0.4]{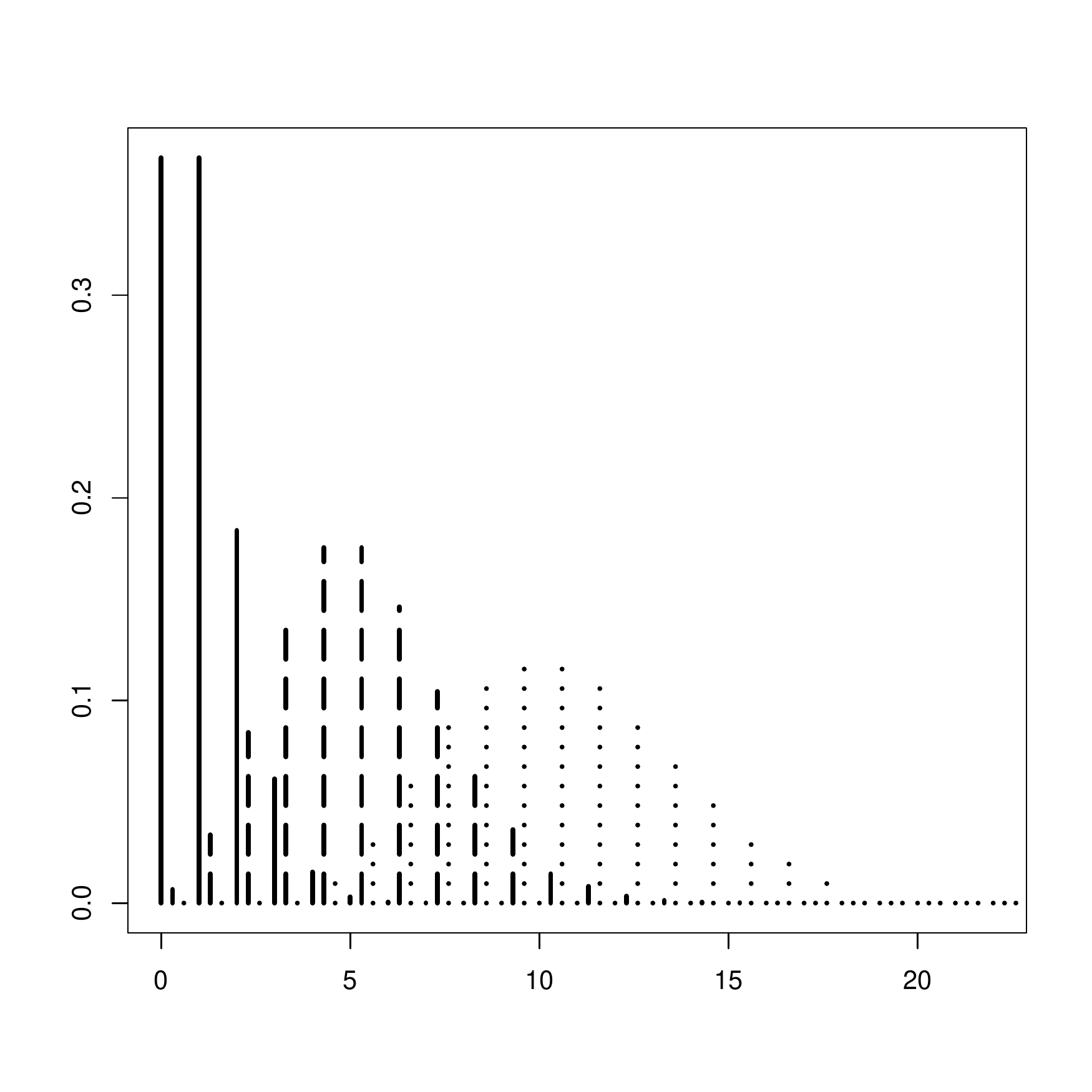}}}
	{\subfigure[Circular densities]{\includegraphics[scale=0.4]{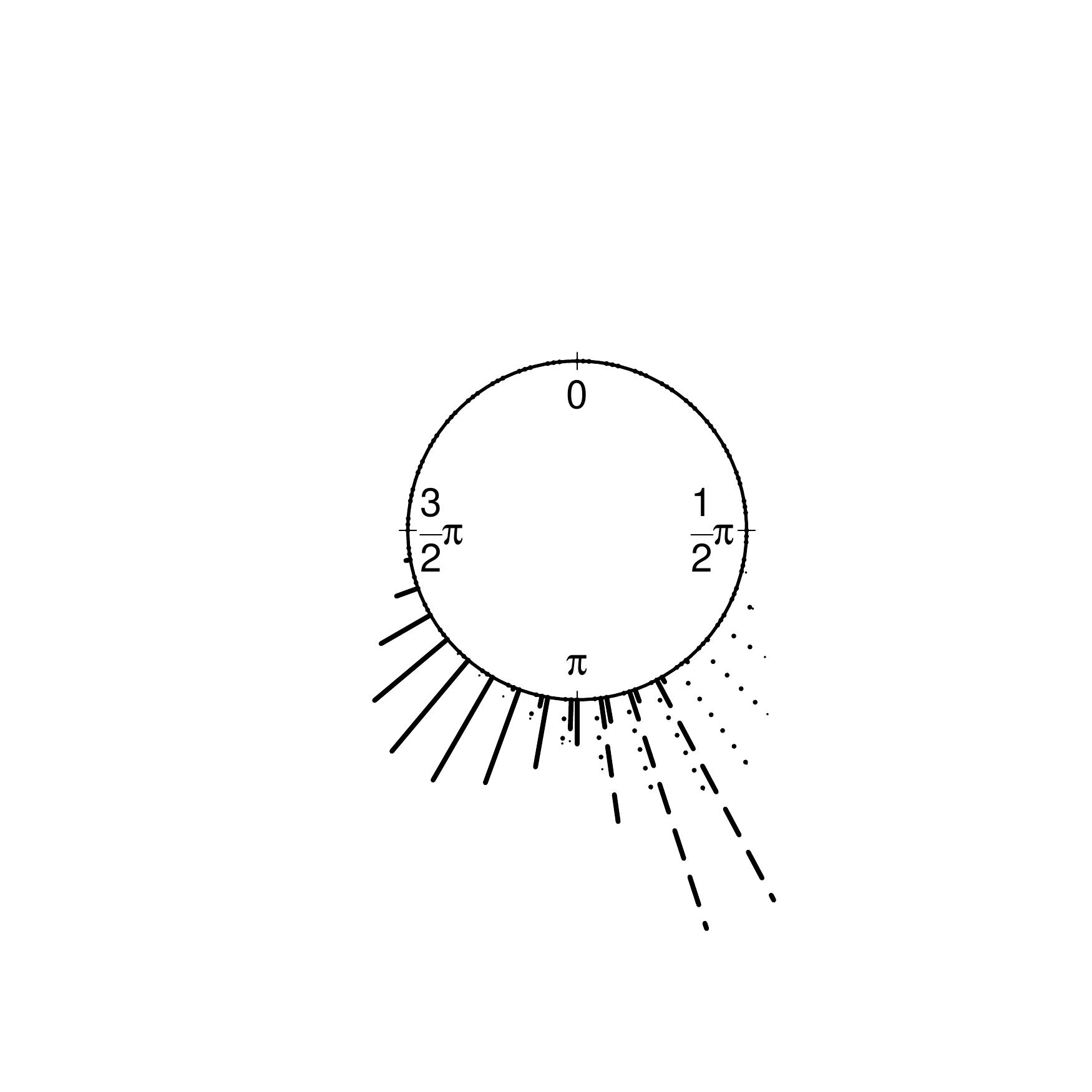}}}
	\caption{Simulate example: densities used to generate the example 2. The solid line is the first regime, the dashed the second and the dotted the third } \label{fig:Ex2}
\end{figure}
\begin{figure}[t!]
	\centering
	{\subfigure[Linear densities]{\includegraphics[scale=0.4]{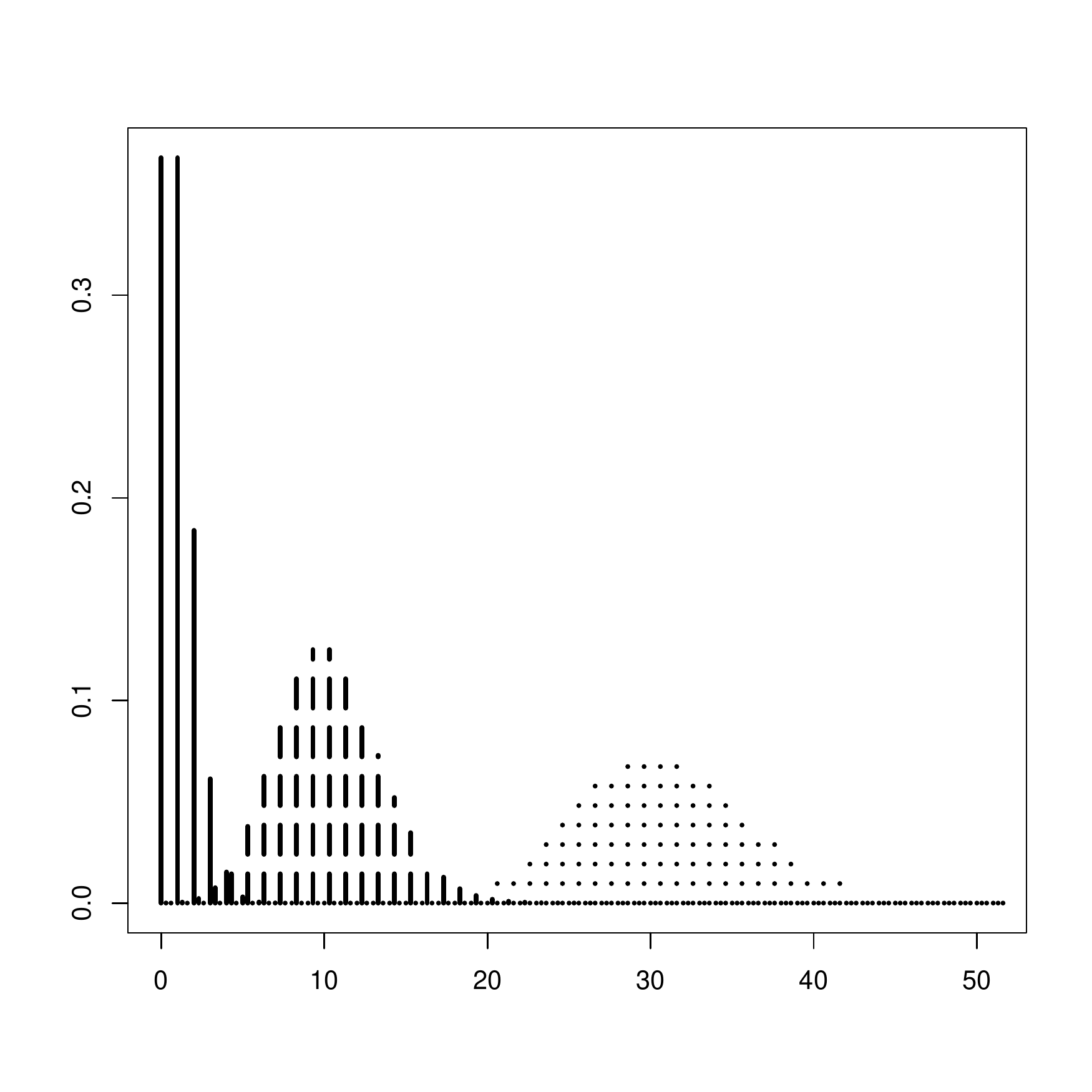}}}
	{\subfigure[Circular densities]{\includegraphics[scale=0.4]{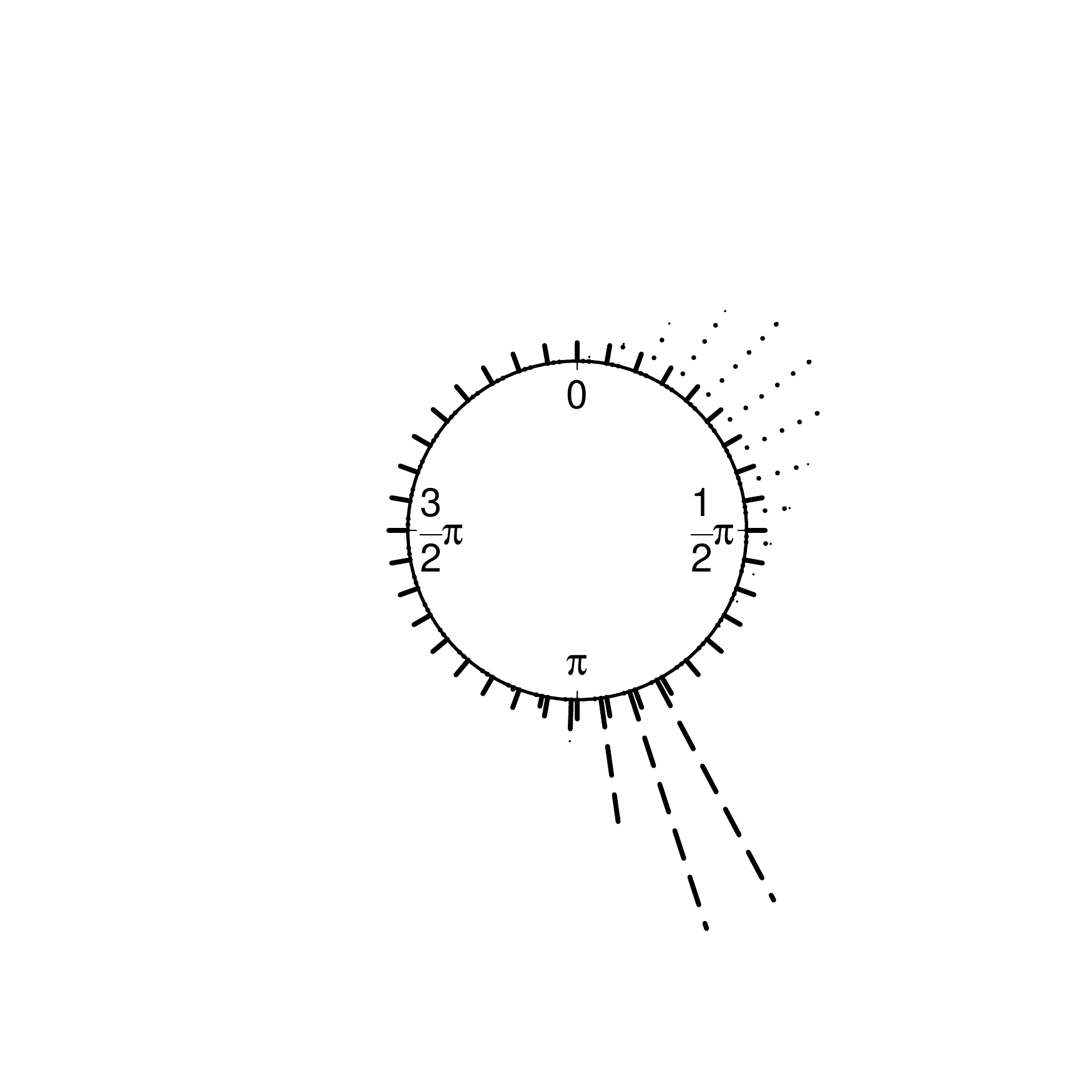}}}
	\caption{Simulate example: densities used to generate the example 3. The solid line is the first regime, the dashed the second and the dotted the third } \label{fig:Ex3}
\end{figure}
\begin{figure}[t!]
	\centering
	{\subfigure[Linear densities]{\includegraphics[scale=0.4]{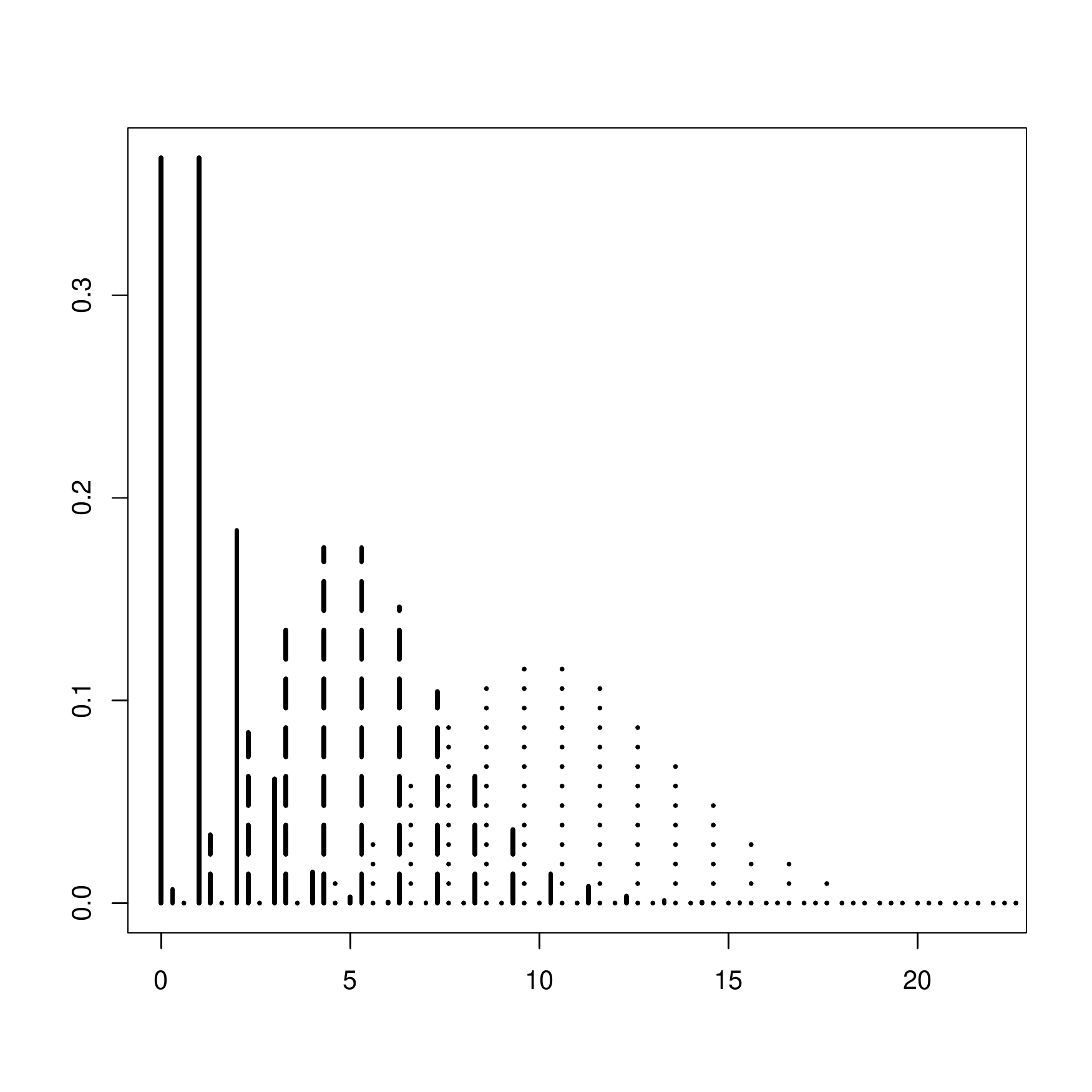}}}
	{\subfigure[Circular densities]{\includegraphics[scale=0.4]{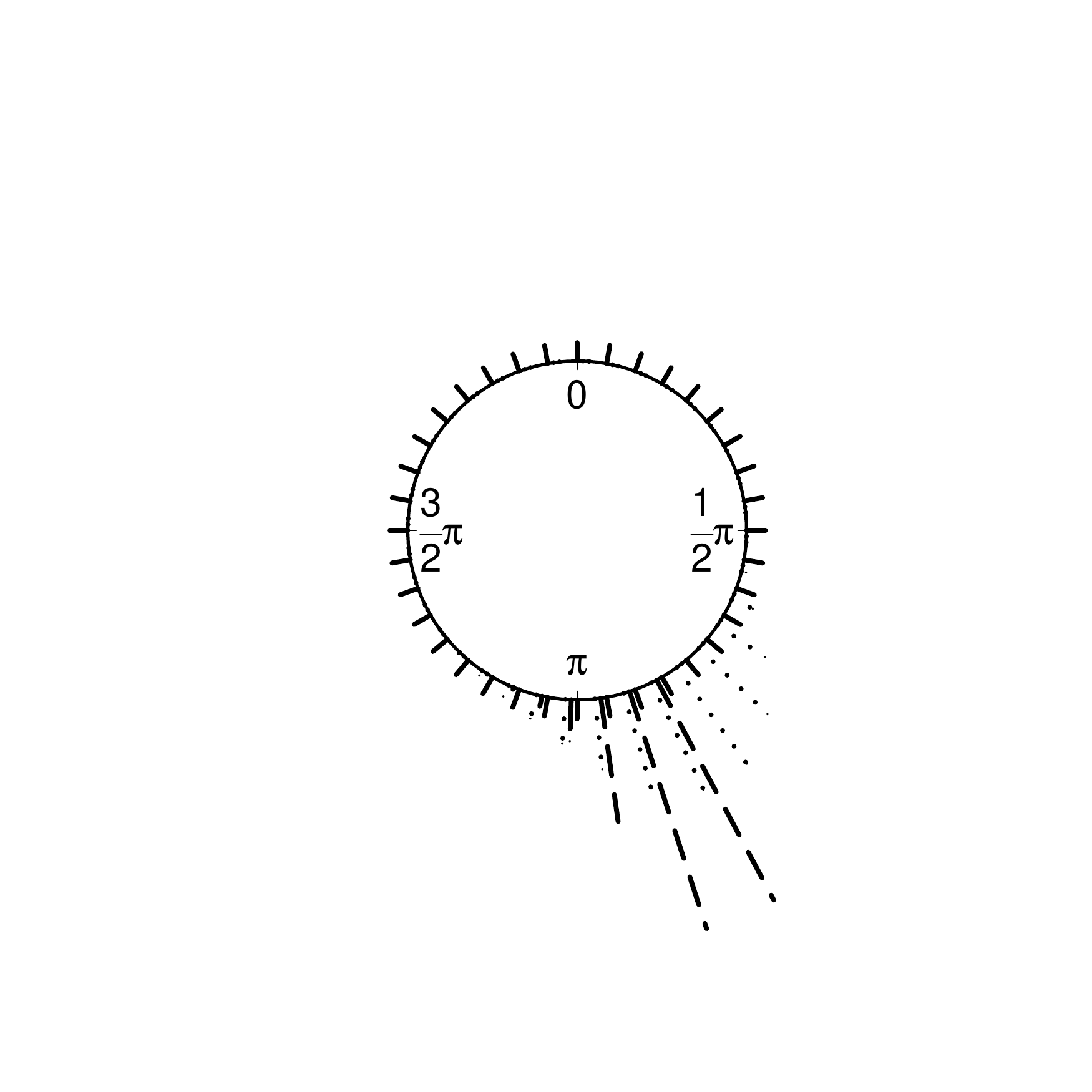}}}
	\caption{Simulate example: densities used to generate the example 4. The solid line is the first regime, the dashed the second and the dotted the third } \label{fig:Ex4}
\end{figure}

\begin{table}[t!]
	\centering
	\begin{tabular}{c|ccc} 
		 & Regime 1 & Regime 2 & Regime 3  \\ \hline \hline
		$\hat{\lambda}_{y,r}$ & 1.056   & 10.059 &  30.103 \\
		CI    &  (0.981 1.138)&  (9.860  10.266)&  (29.771  30.467)  \\
		$\hat{\lambda}_{x,r}$ & 4.887  &  1.024&  5.055\\
		CI    &   (4.740  5.049)&  (0.961 1.093)&  (4.917  5.190)   \\
		$\hat{\nu}_r$ &  0.102  &  0.001& 0.002\\
		CI    &   (0.078  0.124)&   (0.000 0.004)&   (0.000  0.005)    \\ \hline
	\end{tabular}
	\caption{Simulated example: posterior estimates $(\hat{})$ and 95\% credible intervals (CI) for $\lambda_{y,r}$, $\lambda_{x,r}$ and $\nu_{r}$: example 1.} \label{tab:sim1}
\end{table}

\begin{table}[t!]
	\centering
	\begin{tabular}{c|ccc}
		& Regime 1 & Regime 2 & Regime 3  \\ \hline \hline
		$\hat{\lambda}_{y,r}$ &  1.010 & 5.130&  10.249 \\
		CI    &  (0.930  1.093)&  (4.954  5.307)&  (9.993  10.518)  \\
		$\hat{\lambda}_{x,r}$ & 4.931  & 2.075&  4.941\\
		CI    &   (4.765  5.100)&  (1.978  2.173)&  (4.767  5.106)   \\
		$\hat{\nu}_r$ &   0.094 &0.001 & 0.001\\
		CI    &   (0.083 0 0.106)&   (0.000  0.004)&   (0.000  0.004)    \\ \hline
	\end{tabular}
	\caption{Simulated example: posterior estimates $(\hat{})$ and 95\% credible intervals (CI) for $\lambda_{y,r}$, $\lambda_{x,r}$ and $\nu_{r}$: example 2.} \label{tab:sim2}
\end{table}

\begin{table}[t!]
	\centering
	\begin{tabular}{c|ccc}
		& Regime 1 & Regime 2 & Regime 3  \\ \hline \hline
		$\hat{\lambda}_{y,r}$ &  1.015  & 10.029&   30.419\\
		CI    &  (0.940 1.089)&  (9.817  10.244)&  (30.075  30.776)  \\
		$\hat{\lambda}_{x,r}$ &  243.205 & 1.016& 5.041 \\
		CI    &   (170.534  303.098)&  (0.949  1.083)&  (4.896  5.189)   \\
		$\hat{\nu}_r$ &  0.094  &0.001 & 0.001\\
		CI    &   (0.083  0.105)&   (0.000  0.004)&   (0.000  0.004)    \\ \hline
	\end{tabular}
	\caption{Simulated example: posterior estimates $(\hat{})$ and 95\% credible intervals (CI) for $\lambda_{y,r}$, $\lambda_{x,r}$ and $\nu_{r}$: example 3.} \label{tab:sim3}
\end{table}

\begin{table}[t!]
	\centering
	\begin{tabular}{c|ccc}
		& Regime 1 & Regime 2 & Regime 3  \\ \hline \hline
		$\hat{\lambda}_{y,r}$ &  0.954 & 4.907&   10.117\\
		CI    &  (0.881  1.030)&  (4.701  5.111)&  (9.868  10.379)  \\
		$\hat{\lambda}_{x,r}$ &  256.186 &2.927 & 4.919 \\
		CI    &   (183.894  301.445)&  (2.811  3.050)&  (4.759  5.080)   \\
		$\hat{\nu}_r$ &  0.094   & 0.001& 0.001\\
		CI    &   (0.083 0.107)&   (0.000  0.004)&   (0.000  0.004)    \\ \hline
	\end{tabular}
	\caption{Simulated example: posterior estimates $(\hat{})$ and 95\% credible intervals (CI) for $\lambda_{y,r}$, $\lambda_{x,r}$ and $\nu_{r}$:   example 4.} \label{tab:sim4}
\end{table}

\begin{table}[t!]
	\centering
	\begin{tabular}{cc|ccc}
		&&&Destination  \\
		&& Regime 1&Regime 2&Regime 3 \\
		\hline
		&Regime 1& 0.771 & 0.101&0.128 \\
		&& (0.743  0.812)& (0.083  0.121)& (0.098  0.150)\\
Origin 	&	Regime 2& 0.108 & 0.788 & 0.104\\
	&	& (0.090  0.128)& (0.763 0.813)& (0.085  0.122)\\
	&	Regime 3& 0.095 &0.107 & 0.798\\
	&	& (0.079  0.113)& (0.090  0.126)& (0.774  0.820)\\
		\hline
		\hline
	\end{tabular}
	\caption{Simulated example: posterior mean estimates and 95 \% credible intervals for the transition probability matrix: example 1.}  \label{tab:sim5}
\end{table}

\begin{table}[t!]
	\centering
	\begin{tabular}{cc|ccc}
		&&&Destination  \\
		&& Regime 1&Regime 2&Regime 3 \\
		\hline
	&	Regime 1& 0.772 & 0.121& 0.107\\
	&	& (0.742  0.809)& (0.097  0.147)& (0.086  0.132)\\
Origin	&	Regime 2& 0.099 &0.790 &0.110 \\
	&	& (0.081  0.119)& (0.762  0.817)& (0.090  0.135)\\
	&	Regime 3&  0.097& 0.132 & 0.791\\
	&	& (0.077  0.120)& (0.094 0.164)& (0.746  0.820)\\
		\hline
		\hline
	\end{tabular}
	\caption{Simulated example: posterior mean estimates and 95 \% credible intervals for the transition probability matrix: example 2. }  \label{tab:sim6}
\end{table}

\begin{table}[t!]
	\centering
	\begin{tabular}{cc|ccc}
		&&&Destination  \\
		&& Regime 1&Regime 2&Regime 3 \\
		\hline
	&	Regime 1&0.777  & 0.108& 0.114\\
	&	& (0.752  0.802)& (0.090  0.128)& (0.095  0.134)\\
Origin	&	Regime 2&  0.122& 0.774&0.104 \\
	&	& (0.092  0.142)& (0.749  0.800)& (0.085  0.123)\\
	&	Regime 3&  0.107&0.102 & 0.791\\
	&	& (0.090  0.126)& (0.085  0.121)& (0.766  0.813)\\
		\hline
		\hline
	\end{tabular}
	\caption{Simulated example: posterior mean estimates and 95 \% credible intervals for the transition probability matrix: example 3. }  \label{tab:sim7}
\end{table}

\begin{table}[t!]
	\centering
	\begin{tabular}{cc|ccc}
		&&&Destination  \\
		&& Regime 1&Regime 2&Regime 3 \\
		\hline
	&	Regime 1&  0.786&0.112 & 0.101 \\
	&	& (0.760  0.812)& (0.091  0.136)& (0.082 0.123)\\
Origin	&	Regime 2& 0.112 &0.799 & 0.090\\
	&	& (0.091  0.133)& (0.770  0.826)& (0.069  0.113)\\
	&	Regime 3& 0.104 & 0.095&0.800 \\
	&	& (0.084  0.126)& (0.073  0.121)& (0.771  0.828)\\
		\hline
		\hline
	\end{tabular}
	\caption{Simulated example: posterior mean estimates and 95 \% credible intervals for the transition probability matrix:   example 4. }  \label{tab:sim8}
\end{table}

%

\section{Application} \label{sec:ex}

In this Section, the HMM is applied to  simulated  and real data examples.

%

\subsection{Simulated examples} \label{sec:sim}

We simulate 4 datasets with 3 regimes ($R=3$),  from the model described in Section \ref{sec:model}.
The parameters are chosen so that in two examples  (examples 1 and 3) the marginal circular distributions are slightly overlapping (Figures  \ref{fig:Ex1} (b) and \ref{fig:Ex3} (b)) as well as the  linear ones (Figures \ref{fig:Ex1} (a) and \ref{fig:Ex3} (a)), and in two examples  (examples 2 and 4) the  overlapping involves  larger  portions of the  distributions (Figures  \ref{fig:Ex2} and \ref{fig:Ex4}). In the examples 3 and 4, one of the marginal circular distribution is a discrete  uniform. In all the examples we set  $T=3000$ and $\mathbb{D}= \left \{ \frac{2\pi}{36}j \right\}_{j=0}^{35}$. The transition matrix has  diagonal elements equal to 0.8 while the off-diagonal ones are  0.1. \\
We use the following set of parameters:
\begin{itemize}
	\item Example 1:
	\begin{equation}
	\boldsymbol{\lambda}_y= \left[
	\begin{array}{c}
	1 \\10 \\30
	\end{array}
	\right],\,
	\boldsymbol{\lambda}_x= \left[
	\begin{array}{c}
	5 \\1 \\5
	\end{array}
	\right],\,
	\boldsymbol{\eta}= \left[
	\begin{array}{c}
	-1 \\1 \\1
	\end{array}
	\right],\,
	\boldsymbol{\xi}= \left[
	\begin{array}{c}
	5\frac{2\pi}{36} \\15\frac{2\pi}{36} \\0
	\end{array}
	\right],\,
	\boldsymbol{\nu}= \left[
	\begin{array}{c}
	0.1 \\0 \\0
	\end{array}
	\right].
	\end{equation}
	\item Example 2:
	\begin{equation}
	\boldsymbol{\lambda}_y= \left[
	\begin{array}{c}
	1 \\5 \\10
	\end{array}
	\right],\,
	\boldsymbol{\lambda}_x= \left[
	\begin{array}{c}
	5 \\1 \\5
	\end{array}
	\right],\,
	\boldsymbol{\eta}= \left[
	\begin{array}{c}
	-1 \\1 \\1
	\end{array}
	\right],\,
	\boldsymbol{\xi}= \left[
	\begin{array}{c}
	10\frac{2\pi}{36} \\15\frac{2\pi}{36} \\10\frac{2\pi}{36}
	\end{array}
	\right],\,
	\boldsymbol{\nu}= \left[
	\begin{array}{c}
	0.1 \\0 \\0
	\end{array}
	\right].
	\end{equation}
	\item Example 3:
	\begin{equation}
	\boldsymbol{\lambda}_y= \left[
	\begin{array}{c}
	1 \\10 \\30
	\end{array}
	\right],\,
	\boldsymbol{\lambda}_x= \left[
	\begin{array}{c}
	300 \\1 \\5
	\end{array}
	\right],\,
	\boldsymbol{\eta}= \left[
	\begin{array}{c}
	-1 \\1 \\1
	\end{array}
	\right],\,
	\boldsymbol{\xi}= \left[
	\begin{array}{c}
	5\frac{2\pi}{36} \\15\frac{2\pi}{36} \\0
	\end{array}
	\right],\,
	\boldsymbol{\nu}= \left[
	\begin{array}{c}
	0.1 \\0 \\0
	\end{array}
	\right].
	\end{equation}
	\item Example 4;
	\begin{equation}
	\boldsymbol{\lambda}_y= \left[
	\begin{array}{c}
	1 \\5 \\10
	\end{array}
	\right],\,
	\boldsymbol{\lambda}_x= \left[
	\begin{array}{c}
	300 \\1 \\5
	\end{array}
	\right],\,
	\boldsymbol{\eta}= \left[
	\begin{array}{c}
	-1 \\1 \\1
	\end{array}
	\right],\,
	\boldsymbol{\xi}= \left[
	\begin{array}{c}
	10\frac{2\pi}{36} \\15\frac{2\pi}{36} \\10\frac{2\pi}{36}
	\end{array}
	\right],\,
	\boldsymbol{\nu}= \left[
	\begin{array}{c}
	0.1 \\0 \\0
	\end{array}
	\right].
	\end{equation}
\end{itemize}
The  example 4 is particularly challenging since there is a strong overlap between  the circular distributions (see Figure  \ref{fig:Ex4}). \\
Here, and in the real data example of Section \ref{sec:real}, we run  models estimations  using a MCMCs with 100000 iterations, a burnin of 50000 and keep  for inference one observation every  10 samples, i.e. for posterior estimates we use 5000 samples.  Since in the simulated examples and in the  real data one, we never observed a value of the linear variable greater than 50, then  as prior for $\lambda_{y,r}$, the mean of the linear variable, we use  $G(1, 0.00005)I(0, 50)$. 
Following  \cite{mastrantonio2015f}, as  priors for $\lambda_{x,r}$ we use   $G(1, 0.00005)I(0, 500)$. To conclude the priors specification we choose
$\gamma \sim G(1,0.1)$, $\tau \sim G(1,0.1)$, that are standard weak informative distributions, and discrete uniform for $\nu_k$ and $\xi_k$.\\
In all  four examples the posterior  distribution of $R$ is concentrated over 3 (regimes).
The parameters $\lambda_{y,r}$s, $\nu_{r}$s and the transition matrices
are  correctly estimated\footnote{ A parameter is considered correctly estimated if the 95\% credible interval (CI) contains the value used to simulate the data.}, see Tables  from \ref{tab:sim1} to  \ref{tab:sim8}, and the marginal  posterior  distributions of the $\eta_r$s   are   concentrated on the ``true'' values used to simulate  each dataset.\\
In the   examples 1 and 3 both 
$\lambda_{x,r}$ and  $ \xi_r$ are always correctly estimated, except in the first regime of the third example where the posterior distribution of $\xi_1$  has non-zero probability on   22$\frac{2 \pi}{36}$, 23$\frac{2 \pi}{36}$ and 24$\frac{2 \pi}{36}$.
Note that when the density of the discrete circular variable  is  really close to the uniform, the parameters of the IWP becomes weakly identifiable and then we are not surprised that in the third example, the CI of $\lambda_{x,1}$  has length $\approx 133$ and the true value of $\xi_1$ is not inside the associated  CI.  \\
In the  examples 2 and 4, the $\lambda_{x,r}$s are right estimated in the first and third regimes while in the second regime is overestimated. The posterior distributions of the $\xi_r$s are concentrated over the values used to simulate the data in the first and  third  regime of the  example 2 and the first regime of the example 4. In the second regimes of  the example 2 and 4, the posteriors of $\xi_k$ are  concentrated over $14\frac{2\pi}{36}$ and $13\frac{2\pi}{36}$ respectively, while in the third regime,   example 4, is concentrated over 0.\\ 
The model we propose is able to recover the parameters used to simulate  the data in all identifiable  situations.

\subsection{Real data application} \label{sec:real}

\begin{table}[t!]
	\centering
	\begin{tabular}{c|ccc}
		& Regime 1 &Regime 2 & Regime 3 \\ \hline \hline
		$\hat{\lambda}_{y,r}$ & 3.908 & 8.646 &  13.112 \\
		CI    &  (3.794  4.017)&  (8.222  9.061)&  (12.633  13.590)  \\
		$\hat{\lambda}_{x,r}$ & 93.374 & 16.486  & 59.344\\
		CI    &  (92.571  94.145)&  (13.304 20.676 )&  (52.494  60.536)  \\
		$\hat{\nu}_r$& 0.156 &   0.002 & 0.002 \\
		CI    &  (0.141  0.172)&  (0.000 0.009)&  (0.000  0.011)   \\
		$\hat{\mu}_r$& 5.951 &  0.983  & 4.161 \\
		CI    &  (5.817  6.091 )&  (0.885 1.077)&  (3.971 4.347)   \\
		$\hat{c}_r$&  0.242&   0.778 & 0.406 \\
		CI    &  (0.239  0.245)&  (0.730  0.817)&  (0.399  0.450)   \\
		 \hline
	\end{tabular}
	\caption{Real data example: posterior estimates $(\hat{})$ and 95\% credible intervals (CI) for $\lambda_{y,r}$, $\lambda_{x,r}$ and $\nu_{r}$.} \label{tab:eqw}
\end{table}

\begin{table}[t!]
	\centering
	\begin{tabular}{cc|ccc}
	&&&Destination  \\
	&& Regime 1&Regime 2&Regime 3 \\
	\hline
	&	Regime 1&  0.946 &0.037  & 0.017  \\
	&	& (0.934 0.957)& (0.028  0.047)& (0.011 0.025)\\
Origin	&	Regime 2& 0.127 & 0.808 &  0.064\\
	&	& (0.091  0.167)& (0.763  0.850)& (0.039  0.095)\\
	&	Regime 3 & 0.155  & 0.020 &  0.824\\
	&	& (0.115 0.200)& (0.005  0.043)& (0.780  0.867)\\
		\hline
		\hline
	\end{tabular}
	\caption{Real data example:  Posterior mean estimates and 95 \% credible intervals for the transition probability matrix. } \label{tab:traq}
\end{table}

\begin{figure}[t!]
	\centering
	{\subfigure[Linear densities]{\includegraphics[scale=0.4]{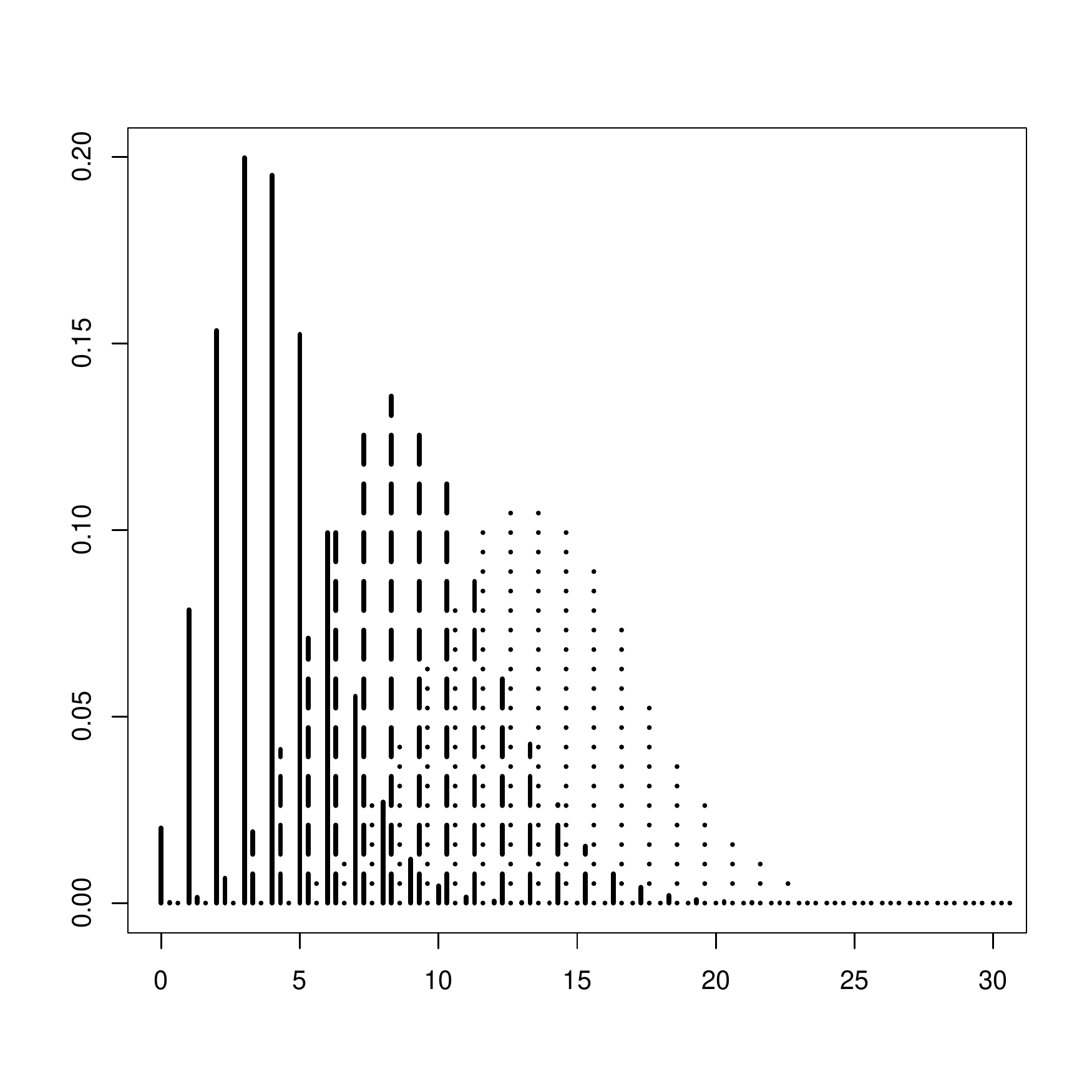}}}
	{\subfigure[Circular densities]{\includegraphics[scale=0.4]{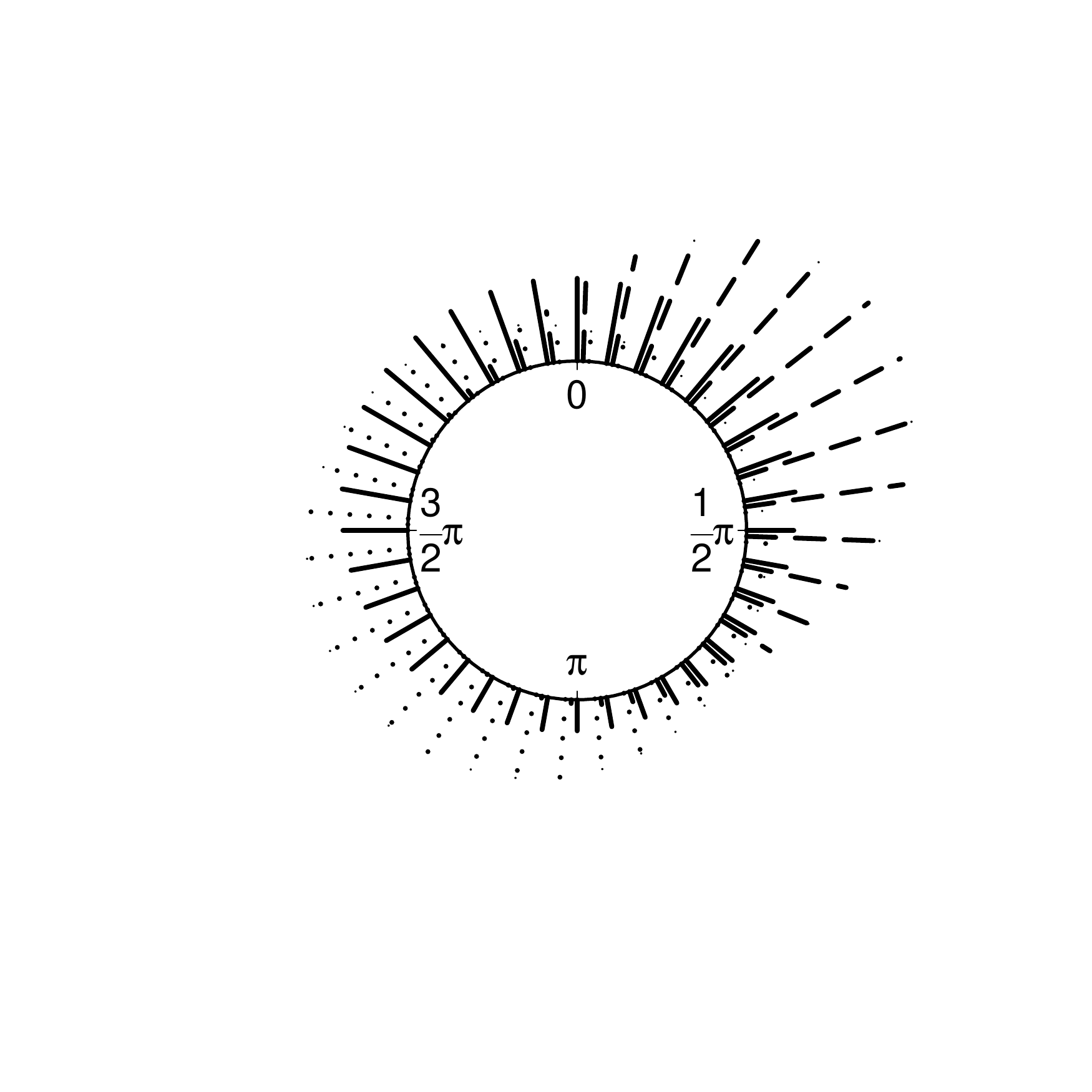}}}
	\caption{Real data example: predictive densities. The solid line is the first regime, the dashed the second and the dotted the third } \label{fig:Ex5}
\end{figure}

%
%
%
%

%

 After the model fitting we observed that 
the posterior distribution of $R$ is concentrated over 3.  The three estimated regimes are numbered in increasing order, based on the value of mean wind speeds, $\lambda_{y,r}$, that can be seen in Table \ref{tab:eqw} along with the posterior estimates of  $\lambda_{x,r}$, $\nu_r$, and the circular mean and concentration. The predictive posterior densities are depicted in Figure   \ref{fig:Ex5}.

With probability 1 the parameter $\eta_r$ is  -1  in the first regime and 1 in the second, while in the third  is equal to -1 with probability 0.989 and 1 with probability 0.011. The posterior distribution of $\xi_r$ is concentrated over $18 \frac{2 \pi}{36}$ in the first regime, in the second regime it has probability greater than 0   between the values  $19 \frac{2 \pi}{36}$ and $31 \frac{2 \pi}{36}$ and its  mode, $27 \frac{2 \pi}{36}$, 
has probability 0.223. In the third regime the posterior distribution of $\xi_r$ assumes positive values between $26 \frac{2 \pi}{36}$ and $6 \frac{2 \pi}{36}$\footnote{$\xi$ is a discrete circular variable and then the left end of an interval can be greater than the right one} and the modal value, $26 \frac{2 \pi}{36}$, has probability 0.845.

In the first regime we have a mean wind speed of 3.908, while the mean wind direction is  5.951, that corresponds to a direction between North and North-West. The circular concentration is 0.242 and the distribution  is  close to the discrete uniform, see Figure \ref{fig:Ex5}. The posterior mean value of  $\nu_r$, the hurdle probability, is 0.156.  According to the empirical Beaufort scale (\cite{scoot2005}),  the first  regime represents the \emph{light breeze} state, where there are ripples without crests or small wavelets and then there is not cost erosion.

In the second regime, that can be considered as a transition state between the calm (first regime) and the storm (third) regime,   the mean wind speed is  8.646 and the mean wind direction is 0.983, roughly North-East, while the circular concentration is 0.778, i.e. the second regime has a directional distribution more concentrated. The hurdle probability is 0.002, really close to zero.

In the third regime the mean wind speed is 13.112 and the mean direction is about South-West, $\hat{\mu}_{3}=4.161$. The circular concentration is 0.406 and again the mean hurdle probability is 0.002. 
 In this regime, the distribution of the wind speed  is fully concentrated between   5 and 23 knots,    resulting in an extreme wave height of almost 4 meters in open water.   
It is  interesting to note that, in this year, the  winds  with the higher speed   are the ones blowing from the sea, more precisely from South West quadrant, resulting in a year with waves with more energy, intensification of erosion and changes in the longshore drift. \\
The posterior distribution of the transition probability matrix, Table \ref{tab:traq}, shows a strong self transition, i.e. the left side interval of the CIs of the self transition are always higher than 0.75.

\section{Discussion} \label{sec:disc}

Motivated by our real data example, we introduced a new HMM  for discrete circular-linear variables. Our data have some peculiar features: i) the linear and circular observations are interval-censored, ii) measurements of wind speed equal to 0 and 1  are not reliable, iii) some of the missing observations of the circular  variables are informative on the values of the non reliable wind speed measurements. All these features was taken  into account when the regime-specific density of the HMM was specified. We introduced a new circular-linear distribution that is suited to model our data. We estimated the model in a non-parametric Bayesian framework and we have shown how specific choices of prior distributions lead to a MCMC algorithm based only on Gibbs steps. We estimated the model on 4 simulated examples and then on the real one.

Future work will find us enriching the model in at least two directions.
First we want to introduce a time-dependent transition matrix. Second, since the Poisson, due to the   unit variance-to-mean ratio, can not model over-dispersed data,  we are going to change the regime-specific marginal linear  density  to increase flexibility.

\bibliographystyle{gSCS}
\bibliography{all}

\end{document}